\documentclass[twoside]{dis04}
\def\runauthor{Benjamin Portheault}
\def\shorttitle{H1 QCD analysis of inclusive cross section data}
\begin{document}

\title{H1 QCD analysis of inclusive cross section data}

\author{BENJAMIN~PORTHEAULT}

\address{On behalf of the H1 collaboration \\
Laboratoire de l'Acc\'el\'erateur Lin\'eaire \\
IN2P3-CNRS et Universit\'e de Paris Sud,\\ 
F-91898 Orsay Cedex\\ 
E-mail: portheau@lal.in2p3.fr}

\maketitle

\abstracts{This contribution reviews the QCD analysis of the H1 inclusive DIS data. Flavor separated parton densities are extracted in a NLO QCD fit  using neutral and charged current  HERA I data. The results of a dedicated QCD analysis of the gluon and $\alpha_s$ are also reviewed.}

At leading order in the electroweak interaction, the double differential cross section of inclusive Deep Inelastic Scattering (DIS) can be expressed in terms of structure functions

\begin{equation}
\frac{\mathrm{d}^2\sigma_{NC}^{\pm}}{\mathrm{d}x\mathrm{d}Q^2}=\frac{2\pi \alpha^2}{xQ^{4}}\left[ Y_{+}\tilde{F}_{2}-y^2\tilde{F}_{L}\mp Y_{-}x\tilde{F}_{3} \right],
\label{NC}
\end{equation}
for Neutral Currents (NC) where $Y_{\pm}=1\pm (1-y)^2$, and similarly for Charged Current (CC)
\begin{equation}
\frac{\mathrm{d}^2\sigma_{CC}^{\pm}}{\mathrm{d}x\mathrm{d}Q^2}=\frac{G_{F}^{2}}{4\pi x}\left[\frac{M_{W}^{2}}{Q^2+M_{W}^{2}} \right]^2 \left[ Y_{+} F_{2}^{CC\pm}-y^{2}F_{L}^{CC\pm}\mp Y_{-}x F_{3}^{CC\pm} \right],
\label{CC}
\end{equation}
where the structure functions exhibit a dependency upon the incoming lepton charge. The QCD factorization theorem allows the separation of the long distance physics and the short distance physics, such that the structure functions can be expressed as convolutions of universal parton distributions (pdfs) and perturbatively computable kernels.  The QCD analysis (the so-called ``QCD fits'') aims at extracting  the pdfs through the  QCD evolution.
It is also possible to  extract  any parameter entering in the expression of the cross section, such as the strong coupling constant $\alpha_{s}$.

\section{The H1 inclusive cross section measurement}

The published H1 data from the HERA I data taking covers a large range in $x$ and $Q^2$. The inclusive NC DIS cross sections are measured to an accuracy of 1--2 \% for statistical uncertainty and 2--3 \% of systematic uncertainty. This gives a stringent constraint on $F_2\propto 4(u+\bar{u})+d+\bar{d}$ over the low and medium $x$ regions. Large $x$ and $Q^2$ data are sensitive to the valence densities, but further statistics are needed to fully exploit this constraint. The wide $Q^2$ range covered allow to constraint $\alpha_s g$ from scaling violations. For the inclusive CC cross section, constant progress in understanding the detectors helped to reduce the systematic uncertainties to about 6 \% but the data are still limited by the statistical error  at large $x$ and $Q^2$. One can define a reduced CC cross section 
\begin{equation}
\tilde{\sigma}_{CC}^{\pm}=\frac{2\pi x}{G_{F}^{2}}\left[\frac{Q^{2}+M_{W}^{2}}{M_{W}^{2}}\right]^2\frac{\mathrm{d}^2\sigma_{CC}^{\pm}}{\mathrm{d}x\mathrm{d}Q^2}
\end{equation}
which reads at the lowest order 
\begin{equation}
 \tilde{\sigma}_{CC}^{+}=x\left[ \bar{u}+\bar{c}+(1-y)^2(d+s) \right]\mbox{ and } \tilde{\sigma}_{CC}^{-}=x\left[ u+c+(1-y)^2(\bar{d}+\bar{s}) \right].
\end{equation}
The measured $\tilde{\sigma}_{CC}^{-}$ and $\tilde{\sigma}_{CC}^{+}$ provide thus a   unique constraint on large $x$ for $u$ and $d$ densities, respectively.

\section{QCD analysis and extraction of parton densities}
\begin{figure}[thb]
\vspace*{12.0cm}
\begin{center}
\includegraphics{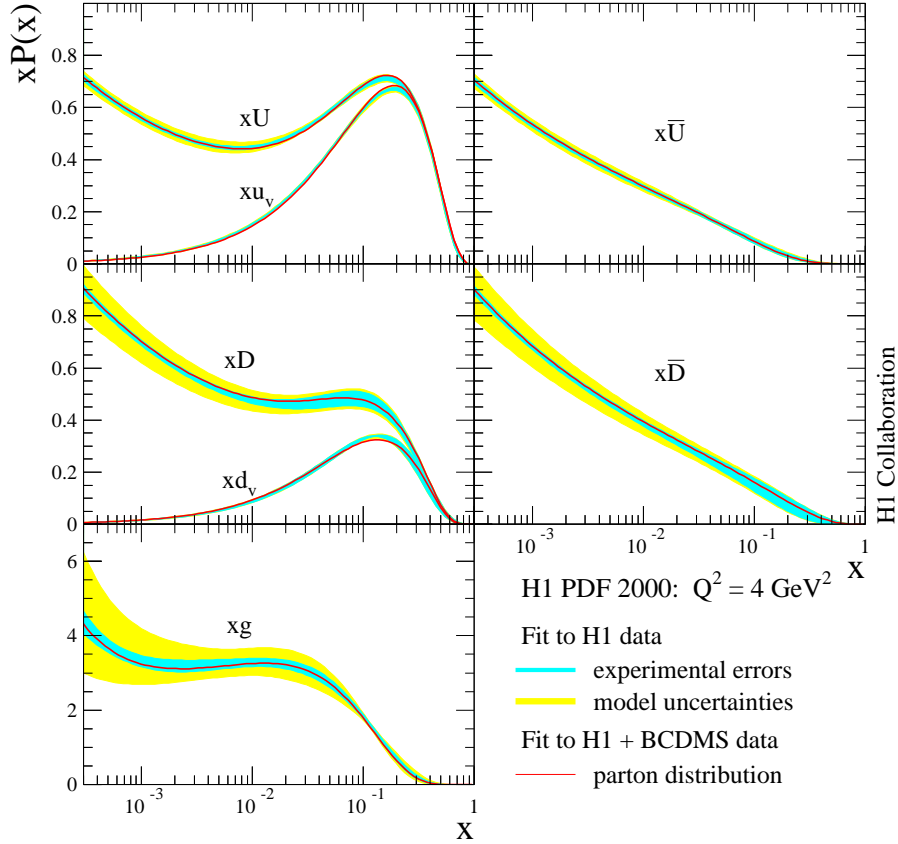}
\caption[*]{\label{fig:pdfs}Parton distributions $xU$, $x\overline{U}$, $xD$, $x\overline{D}$ and $xg$ as determined from the H1PDF2000 fit to H1 data only. The inner error band represents the experimental uncertainty, the outer error band shows the total uncertainty obtained by adding in quadrature the experimental and model uncertainty. For comparison, the parton distributions from the fit to H1 $ep$ and BCDMS $\mu p$ and $\mu d$ data are shown as solid line.}
\end{center}
\end{figure}

All the H1 high $Q^2$ NC and CC data \cite{H1NcCcEp,H1NcCcEm,H1NcCcQCDfit} together with the low $Q^2$ data \cite{H1lowq9697} are used to extract flavor separated parton densities in a NLO QCD analysis \cite{H1NcCcQCDfit} in the so-called massless scheme. The understanding of systematic correlations between the data allows to make reliable error propagation with the Pascaud--Zomer $\chi^2$ method\cite{h1Err}.
To disentangle the different flavors a novel decomposition anzats has been set up which makes use of the flavor combination of the inclusive data: the distributions $U=u+c$ and $D=d+s$ which appear in the NC and CC inclusive cross section with the gluon and the charge conjugate $\overline{U},\overline{D}$ are parameterized at input scale $Q^2_0=4$ GeV$^2$ and evolved. A procedure to explore the parameterization space  has been set up and led to a solution with ten free parameters and $\chi^2/ndof=540/(621-10)=0.88$. As a result of the fit, the $U$ distribution precision is 1\% for $x=0.001$ and 7\% for $x=0.65$. The $D$ distribution precision is 3\% for $x=0.001$ and 30\% for $x=0.65$. Results for the input distributions at $Q^2=4$ GeV$^2$ are shown on Fig. \ref{fig:pdfs}.
A ``model'' uncertainty is estimated by variation of the parameters that enters in the modelisation of the QCD analysis, such as the input scale $Q^2_0$, the strange and charm components of $D$ and $U$, the heavy quark thresholds and the strong coupling constant $\alpha_{s}(M_Z^2)$. Further detail about the analysis can be found in \cite{H1NcCcQCDfit}. 

\section{Determination of the gluon and $\alpha_s$}
\begin{figure}[tbhp]
\vspace*{8.0cm}
\begin{center}
\includegraphics{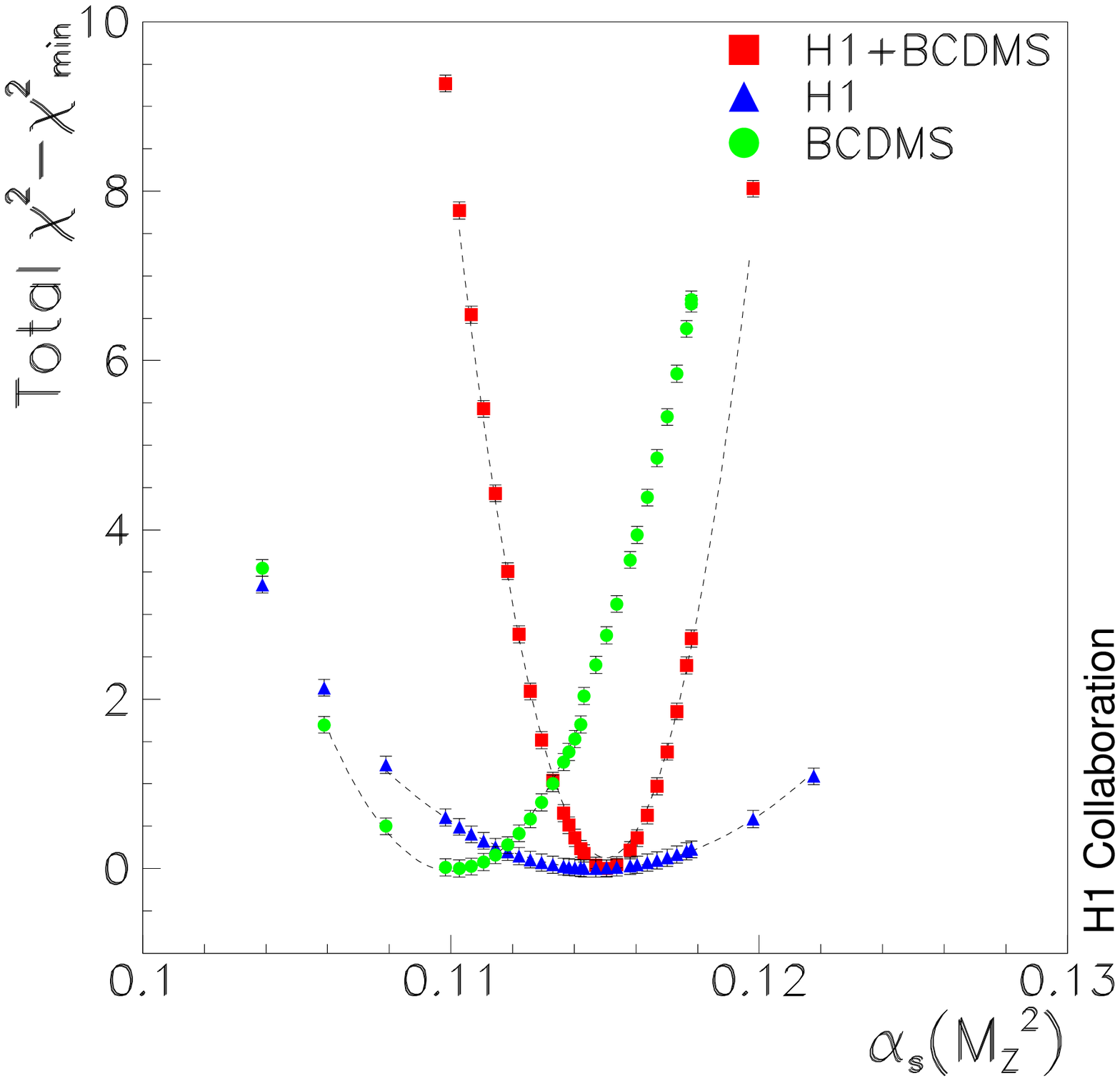}
\includegraphics{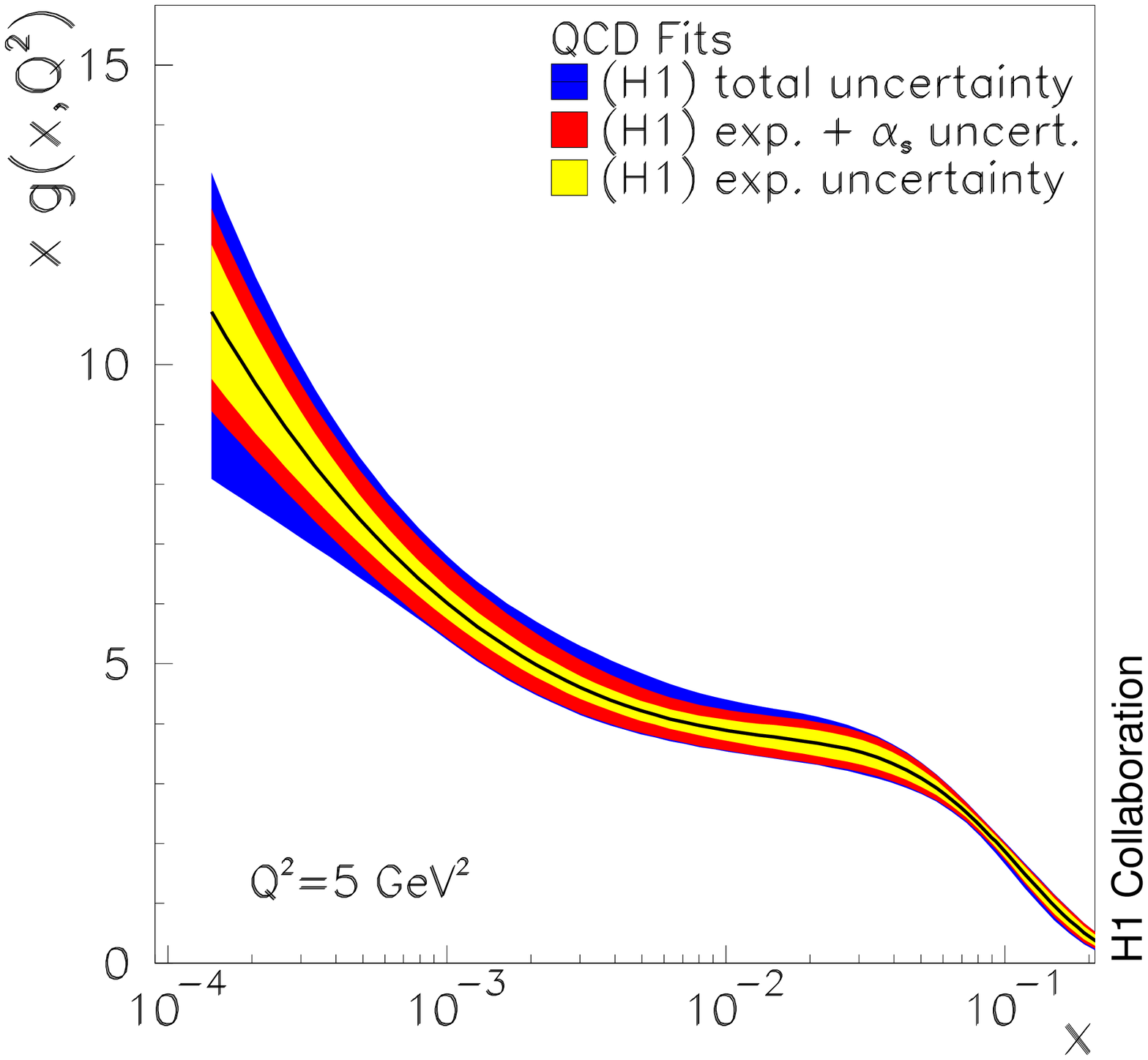}
\caption[*]{\label{fig:alphas}The left plot shows the total $\chi^2$ for the fits of H1 only, BCDMS only and H1+BCDMS fits. The right plot shows the gluon distribution at $Q^2=5$ GeV$^2$. Beside the experimental error, one sees the error due to the $\alpha_s$ uncertainty and the total uncertainty which include the theoretical uncertainty coming from variations of the renormalisation and factorisation scale.}
\end{center}
\end{figure}

Another  QCD analysis \cite{H1lowq9697} dedicated to the determination of the strong coupling constant $\alpha_{s}(M_Z^2)$ and the gluon density was performed in a NLO massive scheme using the H1 low $Q^2$ 96--97 data \cite{H1lowq9697}, the 94--97 high $Q^2$ data \cite{H1NcCcEp} and the BCDMS $\mu p$ DIS data \cite{BCDMSF2p}. The results are shown on Fig. \ref{fig:alphas}.
The result is $\alpha_s=1.1150\pm0.0019(\mbox{exp})\pm0.005(\mbox{th})$ where the theoretical error corresponds to variations of the renormalisation and factorisation scales. The experimental error is small, $\sim 1$\% and the theoretical error is expected to be reduced by factor of $\sim 3$ in a NNLO QCD analysis.

\section{Conclusion}
The determination of flavor separated parton densities form H1 data only and the extraction of the strong coupling $\alpha_s$ and the gluon density are major achievements of the HERA I physics. Further extension to this work -- beside the inclusion of HERA II data or a future NNLO analysis -- could be an inclusion of jet and/or inclusive heavy flavor production data which provide additional constraints on the medium $x$ gluon density.


\begin{thebibliography}{0}

\bibitem{H1NcCcEp}
C.~Adloff {\it et al.}  [H1 Collaboration],
Eur.\ Phys.\ J.\ C {\bf 13}, 609 (2000)
[arXiv:hep-ex/9908059].


\bibitem{H1NcCcEm}
C.~Adloff {\it et al.}  [H1 Collaboration],
Eur.\ Phys.\ J.\ C {\bf 19}, 269 (2001)
[arXiv:hep-ex/0012052].

\bibitem{H1NcCcQCDfit}
C.~Adloff {\it et al.}  [H1 Collaboration],
Eur.\ Phys.\ J.\ C {\bf 30} (2003) 1
[arXiv:hep-ex/0304003].


\bibitem{H1lowq9697}
C.~Adloff {\it et al.}  [H1 Collaboration],
Eur.\ Phys.\ J.\ C {\bf 21} (2001) 33
[arXiv:hep-ex/0012053].


\bibitem{h1Err}
C. Pascaud and F. Zomer, LAL 95-05.

\bibitem{BCDMSF2p}
A.~C.~Benvenuti {\it et al.}  [BCDMS Collaboration],
Phys.\ Lett.\ B {\bf 223}, 485 (1989).

\end{thebibliography}
\end{document}